\documentclass[aps,prl,twocolumn,amsmath,showpacs,superscriptaddress,floatfix,
nopreprintnumbers]{revtex4-1}

\usepackage{verbatim}
\usepackage[T1]{fontenc}
\usepackage[utf8]{inputenc}
\usepackage[american]{babel}
\usepackage{epsfig}
\usepackage{graphicx}
\usepackage{booktabs}
\usepackage{multirow}
\usepackage{dcolumn}
\usepackage{amsmath}
\usepackage{mathtools}
\usepackage{amsfonts}
\usepackage{amssymb}
\usepackage{epstopdf}
\usepackage{bm}
\usepackage{siunitx}
\usepackage{braket}
\usepackage{enumitem}
\usepackage{soul}
\usepackage[table]{xcolor}
\usepackage{color}
\usepackage{transparent}
\usepackage{pifont}
\usepackage[normalem]{ulem}

\definecolor{navyblue}{rgb}{0.0, 0.0, 0.5}
\definecolor{royalblue}{rgb}{0.25, 0.41, 0.88}
\definecolor{cadmiumgreen}{rgb}{0.0, 0.42, 0.24}
\definecolor{blue-violet}{rgb}{0.54, 0.17, 0.89}
\definecolor{darkviolet}{rgb}{0.58, 0.0, 0.83}
\definecolor{orange(colorwheel)}{rgb}{1.0, 0.5, 0.0}

\usepackage{hyperref}
\hypersetup{
    colorlinks=true, 
    linkcolor=royalblue, 
    citecolor=magenta}

\newcommand{\diff}{\mathop{}\!\mathrm{d}}

\usepackage{booktabs}
\usepackage{multirow}
\usepackage{dcolumn}
\usepackage{colortbl}

\definecolor{magenta(process)}{rgb}{1.0, 0.0, 0.56}

\definecolor{darkspringgreen}{rgb}{0.09, 0.45, 0.27}

\definecolor{royalblue(web)}{rgb}{0.25, 0.41, 0.88}

\begin{document}

\title{An absolute $\nu$ mass measurement with the DUNE experiment} 

\author{Federica Pompa}
\email{federica.pompa@ific.uv.es}

\author{Francesco Capozzi}
\email{fcapozzi@ific.uv.es}

\author{Olga Mena}
\email{omena@ific.uv.es}

\author{Michel Sorel}
\email{sorel@ific.uv.es}

\affiliation{Instituto de F\'{i}sica Corpuscular (IFIC), University of Valencia-CSIC, Parc Cient\'{i}fic UV, c/ Cate\-dr\'{a}tico Jos\'{e} Beltr\'{a}n 2, E-46980 Paterna, Spain}

\date{\today}

\preprint{}
\begin{abstract}
Time of flight delay in the supernova neutrino signal offers a unique tool to set model-independent constraints on the absolute neutrino mass. The presence of a sharp time structure during a first emission phase, the so-called neutronization burst in the electron neutrino flavor time distribution, makes this channel a very powerful one. Large liquid argon underground detectors will provide precision measurements of the time dependence of the electron neutrino fluxes. We derive here a new $\nu$ mass sensitivity attainable at the future DUNE far detector from a future supernova collapse in our galactic neighborhood, finding a sub-eV reach under favorable scenarios. These values are competitive with those expected for laboratory direct neutrino mass searches.
\end{abstract}
\maketitle
\section{Introduction} \label{sec:intro} 
Neutrinos of astrophysical and cosmological origin have been crucial for unraveling neutrino masses and properties. Solar neutrinos provided the first evidence for neutrino oscillations, and hence massive neutrinos. We know that at least two massive neutrinos should exist, as required by the two distinct squared mass differences measured, the atmospheric $\lvert\Delta m^2_{31}\rvert \approx 2.51\cdot 10^{-3}$~eV$^2$ and the solar $\Delta m^2_{21} \approx 7.42\cdot 10^{-5}$~eV$^2$ splittings~\cite{deSalas:2020pgw,Esteban:2020cvm,Capozzi:2021fjo}~\footnote{The current ignorance on the sign of $\lvert\Delta m^2_{31}\rvert$ is translated into two possible mass orderings. In the \emph{normal} ordering (NO), the total neutrino mass is $\sum m_\nu \gtrsim 0.06$~eV, while in the \emph{inverted} ordering (IO) it is $\sum m_\nu \gtrsim 0.10 $~eV.}. However, neutrino oscillation experiments are not sensitive to the absolute neutrino mass scale. On the other hand, cosmological observations provide the most constraining recent upper bound on the total neutrino mass via relic neutrinos, $\sum m_\nu<0.09$~eV at $95\%$ confidence level (CL)\cite{DiValentino:2021hoh}, where the sum runs over the distinct neutrino mass states. However, this limit is model-dependent, see for example~\cite{DiValentino:2015sam,Palanque-Delabrouille:2019iyz,Lorenz:2021alz,Poulin:2018zxs,Ivanov:2019pdj,Giare:2020vzo,Yang:2017amu,Vagnozzi:2018jhn,Gariazzo:2018meg,Vagnozzi:2017ovm,Choudhury:2018byy,Choudhury:2018adz,Gerbino:2016sgw,Yang:2020uga,Yang:2020ope,Yang:2020tax,Vagnozzi:2018pwo,Lorenz:2017fgo,Capozzi:2017ipn,DiValentino:2021zxy,DAmico:2019fhj,Colas:2019ret}. 

The detection of supernova (SN) neutrinos can also provide constraints on the neutrino mass, by exploiting the time of flight delay~\cite{Zatsepin:1968ktq} experienced by a neutrino of mass $m_\nu$ and energy $E_\nu$:
\begin{equation}
    \label{eq:delay}
    \Delta t = \frac{D}{2c}\left(\frac{m_\nu}{E_{\nu}}\right)^2~,
\end{equation}

\noindent where $D$ is the distance travelled by the neutrino. This method probes the same neutrino mass constrained via laboratory-based kinematic measurements of beta-decay electrons~\cite{Aker:2021gma,Drexlin:2013lha}. Using neutrinos from SN1987A~\cite{Kamiokande-II:1989hkh,Kamiokande-II:1987idp,Bionta:1987qt,Alekseev:1988gp,Alekseev:1987ej}, a $95\%$ CL current upper limit of $m_\nu<5.8$~eV~\cite{Pagliaroli:2010ik} has been derived (see also \cite{Loredo:2001rx}). Prospects for future SN explosions may reach the sub-eV level~\cite{Pagliaroli:2010ik,Nardi:2003pr,Nardi:2004zg,Lu:2014zma,Hyper-Kamiokande:2018ofw,Hansen:2019giq}. Nevertheless, these forecasted estimates rely on the detection of inverse $\beta$ decay events in water Cherenkov or liquid scintillator detectors, mostly sensitive to $\bar{\nu}_e$ events. An appealing and alternative possibility is the detection of $\nu_e$ exploiting the liquid argon technology at the DUNE far detector~\cite{DUNE:2020zfm,Rossi-Torres:2015rla}. The large number of detected neutrinos and the very distinctive feature of the neutronization burst will ensure a unique sensitivity to the neutrino mass signature via time delays. 

\section{Supernova electron neutrino events} \label{sec:events}
Core-collapse supernovae emit $99\%$ of their energy ($\simeq 10^{53}$~ergs) in the form of neutrinos and antineutrinos of all flavors with mean energies of $\mathcal{O}(10~\si{\mega\electronvolt})$. The explosion mechanism of a core-collapse SN can be divided into three phases: the \emph{neutronization burst}, the \emph{accretion phase} and the \emph{cooling phase}. The first phase, which lasts for 25 milliseconds approximately, is due to a fast \emph{neutronization} of the stellar nucleus via electron capture by free protons, causing an emission of electron neutrinos ($e^- + p\rightarrow \nu_e + n$). The flux of $\nu_e$ stays trapped behind the shock wave until it reaches sufficiently low densities for neutrinos to be suddenly released. Unlike subsequent phases, the neutronization burst phase has little dependence on the progenitor star properties. In numerical simulations, there is a second \emph{accretion} phase of $\sim 0.5$~s in which the shock wave leads to a hot accretion mantle around the high density core of the neutron star. High luminosity $\nu_e$ and $\bar{\nu}_e$ fluxes are radiated via the processes $e^- + p\rightarrow \nu_e + n$ and $e^+ + n \rightarrow \bar{\nu}_e + p$ due to the large number of nucleons and the presence of a quasi-thermal $e^+e^-$ plasma. Finally, in the \emph{cooling} phase, a hot neutron star is formed. This phase is characterized by the emission of fluxes of neutrinos and anti-neutrinos of all species within tens or hundreds of seconds.

For numerical purposes, we shall make use of the following quasi-thermal parametrization, representing well detailed numerical simulations~\cite{Keil:2002in,Hudepohl:2009tyy,Tamborra:2012ac,Mirizzi:2015eza}: 
\begin{equation}
\label{eq:differential_flux}
\Phi^{0}_{\nu_\beta}(t,E) = \frac{L_{\nu_\beta}(t)}{4 \pi D^2}\frac{\varphi_{\nu_\beta}(t,E)}{\langle E_{\nu_\beta}(t)\rangle}\,,
\end{equation}
and describing the differential flux for each neutrino flavor $\nu_\beta$ at a time $t$ after the SN core bounce, located at a distance $D$. In Eq.~\ref{eq:differential_flux}, $L_{\nu_\beta}(t)$ is the $\nu_\beta$ luminosity, $\langle E_{\nu_\beta}(t)\rangle$ the mean neutrino energy and $\varphi_{\nu_\beta}(t,E)$ is the neutrino energy distribution, defined as:
\begin{equation}
\label{eq:nu_energy_distribution}
\varphi_{\nu_\beta}(t,E) = \xi_\beta(t) \left(\frac{E}{\langle E_{\nu_\beta}(t)\rangle}\right)^{\alpha_\beta(t)} \exp{\left\{\frac{-\left[\alpha_\beta(t) + 1\right] E}{\langle E_{\nu_\beta}(t)\rangle}\right\}}\,,
\end{equation}

\noindent where $\alpha_\beta(t)$ is a \emph{pinching} parameter and $\xi_\beta(t)$ is a unit-area normalization factor.

The input for luminosity, mean energy and pinching parameter values have been obtained from the \texttt{SNOwGLoBES} software \cite{snowglobes}. \texttt{SNOwGLoBES} includes fluxes from the Garching Core-Collapse Modeling Group~\footnote{\url{https://wwwmpa.mpa-garching.mpg.de/ccsnarchive/index.html}}, providing simulation results for a progenitor star of $8.8 M_\odot$~\cite{Hudepohl:2009tyy}.

Neutrinos experience flavor conversion inside the SN as a consequence of their coherent interactions with electrons, protons and neutrons in the medium, being subject to the MSW (Mikheyev-Smirnov-Wolfenstein) resonances associated to the solar and atmospheric neutrino sectors~\cite{Dighe:1999bi}. After the resonance regions, the neutrino mass eigenstates travel incoherently on their way to the Earth, where they are detected as flavor eigenstates.  The neutrino fluxes at the Earth ($\Phi_{\nu_e}$ and $\Phi_{\nu_\mu}=\Phi_{\nu_\tau}=\Phi_{\nu_x}$) can be written as:
\begin{eqnarray}
\label{eq:nue}
   \Phi_{\nu_e}&= &p  \Phi^{0}_{\nu_e} +(1-p) \Phi^{0}_{\nu_x}~;\\
    \Phi_{\nu_\mu}+\Phi_{\nu_\tau} \equiv 2\Phi_{\nu_x} & =& (1-p) \Phi^{0}_{\nu_e} + (1+p) \Phi^{0}_{\nu_x}~,
\end{eqnarray}

\noindent where $\Phi^{0}$ refers to the neutrino flux in the SN interior, and the $\nu_e$ survival probability $p$ is given by $p = |U_{e3}|^2= \sin^2 \theta_{13}$ ($p \simeq |U_{e2}|^2 \simeq \sin^2 \theta_{12}$) for NO (IO), due to adiabatic transitions in the $H$ ($L$) resonance, which refer to flavor conversions associated with the atmospheric $\Delta m^2_{31}$ (solar $ \Delta m^2_{21}$) mass splitting, see e.g.~\cite{Dighe:1999bi}. Here we are neglecting possible non-adiabaticity effects occurring when the resonances occur near the shock wave \cite{Schirato:2002tg,Fogli:2003dw,Fogli:2004ff,Tomas:2004gr,Dasgupta:2005wn,Choubey:2006aq,Kneller:2007kg,Friedland:2020ecy}, and the presence of turbulence in the matter density \cite{Fogli:2006xy,Friedland:2006ta,Kneller:2010sc,Lund:2013uta,Loreti:1995ae,Choubey:2007ga,Benatti:2004hn,Kneller:2013ska,Fogli:2006xy}. The presence of non-linear collective effects~\cite{Mirizzi:2015eza,Chakraborty:2016yeg,Horiuchi:2018ofe,Tamborra:2020cul,Capozzi:2022slf} is suppressed by the large flavor asymmetries of the neutronization burst~\cite{Mirizzi:2015eza}.

Earth matter regeneration effects might also affect the neutrino propagation when the SN is shadowed by the Earth. The distance travelled by neutrinos through the Earth depends on a zenith angle $\theta$, analogous to the one usually defined for atmospheric neutrino studies. This convention assumes $\cos \theta=-1$ for neutrinos that cross a distance equal to the Earth's diameter, and $\cos \theta\geq 0$ for neutrinos that are un-shadowed by the Earth. We have implemented such effects using the approach proposed in \cite{Lisi:1997yc} and we verified that they marginally affect the sensitivity to the neutrino mass (see also Tab. \ref{tab:m_nu_mass_bounds}). 

The neutrino interaction rate per unit time and energy in the DUNE far detector is defined as:
\begin{equation}
\label{eq:rate_DUNE_fun}
R(t,E) = N_\text{target}~\sigma_{\nu_e\text{CC}}(E)~\epsilon(E)~\Phi_{\nu_e}(t,E)~,
\end{equation}
\noindent where $t$ is the neutrino emission time, $E$ is the neutrino energy, $N_\text{target}=\num{6.03e32}$ is the number of argon nuclei for a $40$ kton fiducial mass of liquid argon, $\sigma_{\nu_e\text{CC}}(E)$ is the $\nu_e$ cross-section, $\epsilon(E)$ is the DUNE reconstruction efficiency and $\Phi_{\nu_e}(t,E)$ is the electron neutrino flux reaching the detector per unit time and energy. The total number of expected events is given by $R\equiv \int R(t,E)\diff t \diff E$, where $t\in[0,9]$ seconds.

As far as cross-sections are concerned, liquid argon detectors are mainly sensitive to electron neutrinos via their charged-current interactions with $^{40}$Ar nuclei, $\nu_e + {^{40} Ar} \rightarrow e^{-} + {^{40} K^{*}}~$, through the observation of the final state electron plus the de-excitation products (gamma rays, ejected nucleons) from $^{40} K^{*}$. We use the MARLEY~\footnote{MARLEY (Model of Argon Reaction Low Energy Yields), see \url{http://www.marleygen.org/} and \cite{Gardiner:2021qfr}.} charged-current $\nu_e$ cross-section on $^{40}$Ar, implemented in \texttt{SNOwGLoBES} \cite{snowglobes} (see also \cite{Capozzi:2018dat} for a detailed review). Concerning event reconstruction, we assume the efficiency curve as a function of neutrino energy given in \cite{DUNE:2020zfm}, for the most conservative case quoted there of 5~MeV as deposited energy threshold. 

\begin{figure}
\begin{center}
\includegraphics[width=\columnwidth]{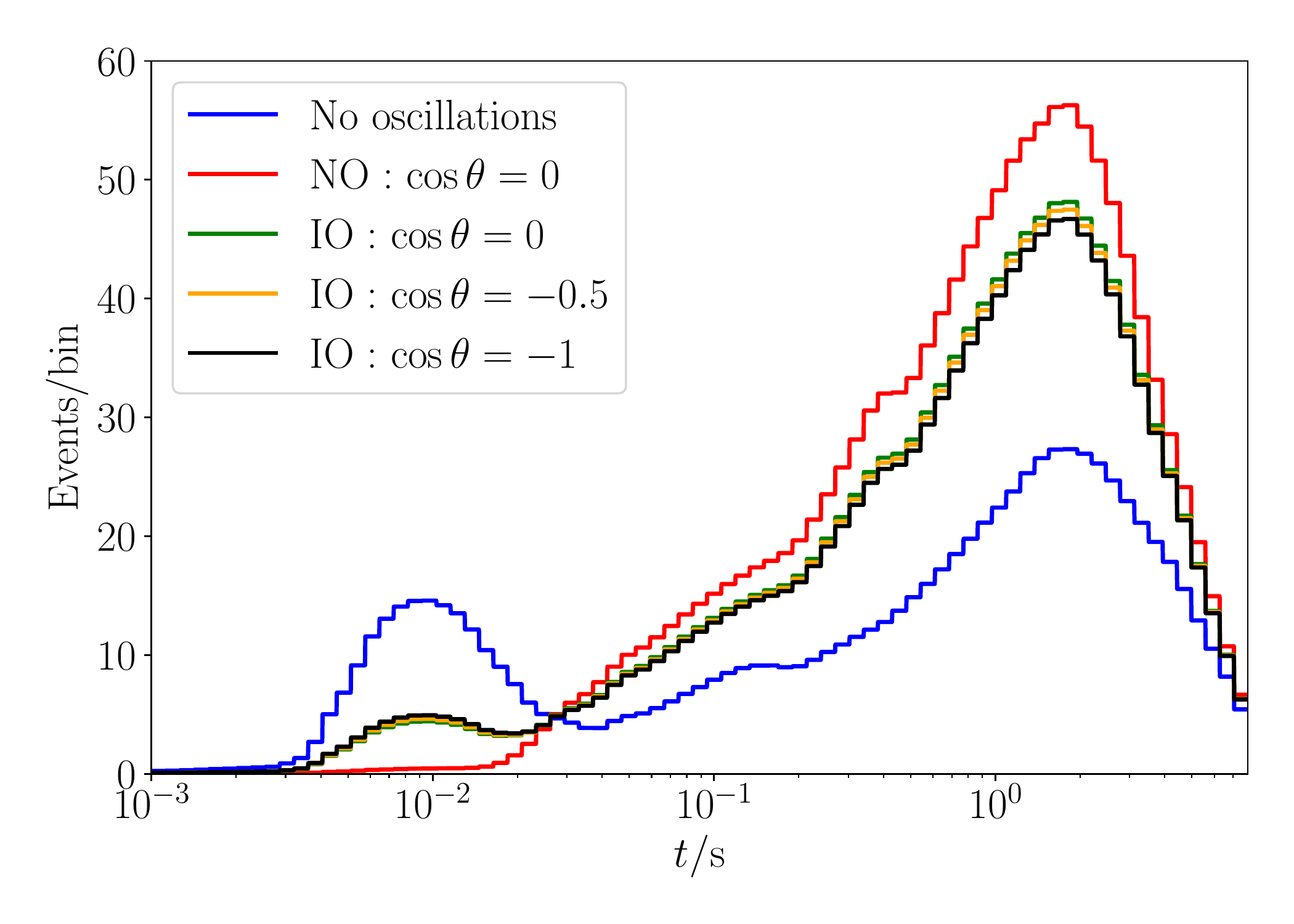} 
\caption{\label{fig:events}Number of $\nu_e$ events as a function of time, 
obtained by an energy integration of Eq.~\ref{eq:rate_DUNE_fun}. 
The number of events per time bin is shown for equal bin widths in logarithmic space, for more clarity. A SN distance of 10~kpc is assumed. Several histograms are shown: neglecting oscillations, and including oscillations for the NO and IO cases. For IO, we show the variation of the Earth matter effects with zenith angle $\theta$.}
\end{center}
\end{figure}

Figure~\ref{fig:events} shows the number of $\nu_e$ events as a function of emission time at the DUNE far detector from a SN explosion at $10$~kpc from Earth, for negligible time delays due to non-zero neutrino masses. Assuming no oscillations, the plot illustrates a clear neutronization burst peak at early times. We also account for oscillations in NO and IO cases, the latter for several possible SN locations with respect to the Earth. The neutronization burst is almost entirely (partially) suppressed for NO (IO).
For a SN located at $D=10$~kpc from the Earth and without Earth matter effects, $R$ is found to be 860, 1372 and 1228 for the no oscillations, NO and IO cases, respectively, among which 201, 54 and 95 come from the first 50~ms. In other words, the largest total event rate is obtained for the largest swap of electron with muon/tau neutrinos in the SN interior, \emph{i.e.} the smallest value of $p$ in Eq.~\ref{eq:nue}, corresponding to the NO case. This can be understood from the larger average neutrino energy at production of muon/tau neutrinos compared to electron neutrinos, resulting in a higher (on average) neutrino cross-section and reconstruction efficiency.

Finally, as shown in Fig.~\ref{fig:events}, Earth matter effects are expected to have a mild effect on the event rate in all cases. The $\nu_e$ flux is left unchanged for NO, while for IO the total number of events becomes $R=1214$ and $1200$ for $\cos\theta = -0.5$ and $-1$, respectively.

\section{Neutrino mass sensitivity} \label{sec:likelihood}
In order to compute the DUNE sensitivity to the neutrino mass, we adopt an unbinned maximum likelihood method similar to the one in \cite{Pagliaroli:2010ik}. However, here we do not include any background or uncertainties on the neutrino production, propagation and interaction. We justify these assumptions in the Supplemental Material.

We start by generating many DUNE toy experiment datasets (a few hundred, typically) for each neutrino oscillation and SN distance scenario, and assuming massless neutrinos. For each dataset, the time/energy information of the $R$ generated events are sampled following the parametrization of Eq.~\ref{eq:rate_DUNE_fun}, and events are sorted in time-ascending order. 
Furthermore, we assume a $10\%$ fractional energy resolution in our $\mathcal{O}$(10~MeV) energy range of interest, see~\cite{DUNE:2020zfm}, and smear the neutrino energy of each generated event accordingly. 
On the other hand, we assume perfect time resolution for our studies. The latter is a good approximation if scintillation light is detected for most SN neutrino interactions, and correctly associated with charge readout information from the TPC. In this case, a time resolution better than 1~$\mu$s is expected \cite{DUNE:2020zfm}, yielding a completely negligible time smearing effect. While detailed studies are still missing, the high light yields expected in the DUNE far detector \cite{DUNE:2020txw} imply that this is a realistic assumption. 

Once events are generated for each DUNE dataset, we proceed with our minimization procedure. The two free parameters constrained in our fit are an offset time $t_\text{off}$ between the moment when the earliest SN burst neutrino reaches the Earth and the detection of the first event $i=1$, and the neutrino mass $m_\nu$. The fitted emission times $t_{i,fit}$ for each event $i$ depend on these two fit parameters as follows:
\begin{equation}
\label{eq:emission_t}
t_{i,fit} = \delta t_i  - \Delta t_{i}(m_\nu) + t_\text{off}\,,
\end{equation}
where $\delta t_i $ is the time at which the neutrino interaction $i$ is measured in DUNE (with the convention that $\delta t_1\equiv 0$ for the first detected event), $\Delta t_i(m_\nu)$ is the delay induced by the non-zero neutrino mass (see Eq.~\ref{eq:delay}), and $t_\text{off}$ is the offset time. 

By neglecting all the constant (irrelevant) factors, our likelihood $\mathcal{L}$ function \cite{Pagliaroli:2008ur} reads as
\begin{equation}
\label{eq:likelihood_fun}
\mathcal{L}(m_{\nu},t_\text{off}) = \prod_{i=1}^{R}\int R(t_i,E_i)G_i(E)\diff E~, 
\end{equation}
\noindent where $G_i$ is a Gaussian distribution with mean $E_i$ and sigma $0.1E_i$, accounting for energy resolution. The estimation of the $m_\nu$ fit parameter is done by marginalizing over the nuisance parameter $t_\text{off}$. For each fixed $m_\nu$ value, we minimize the following $\chi^2$ function:
\begin{equation}
\label{eq:chi2_fun}
\chi^2(m_{\nu}) = -2 \log(\mathcal{L}(m_{\nu},t_\text{off,best}))~,
\end{equation}
\noindent where $\mathcal{L}(m_{\nu},t_\text{off,best})$ indicates the maximum likelihood at this particular $m_\nu$ value. 

The final step in our analysis is the combination of all datasets for the same neutrino oscillation and SN distance scenario, to evaluate the impact of statistical fluctuations. For each $m_\nu$ value, we compute the mean and standard deviation of all toy dataset $\chi^2$ values. In order to estimate the allowed range in $m_\nu$, the $\Delta\chi^2$ difference between all mean $\chi^2$ values and the global mean $\chi^2$ minimum is computed. The mean 95\% CL sensitivity to $m_\nu$ is then defined as the largest $m_\nu$ value satisfying $\Delta \chi^2<3.84$. The $\pm 1\sigma$ uncertainty on the 95\% CL $m_\nu$ sensitivity can be computed similarly, including into the $\Delta\chi^2$ evaluation also the contribution from the standard deviation of all toy dataset $\chi^2$ values.

\begin{figure}
\begin{center}
\includegraphics[width=\columnwidth]{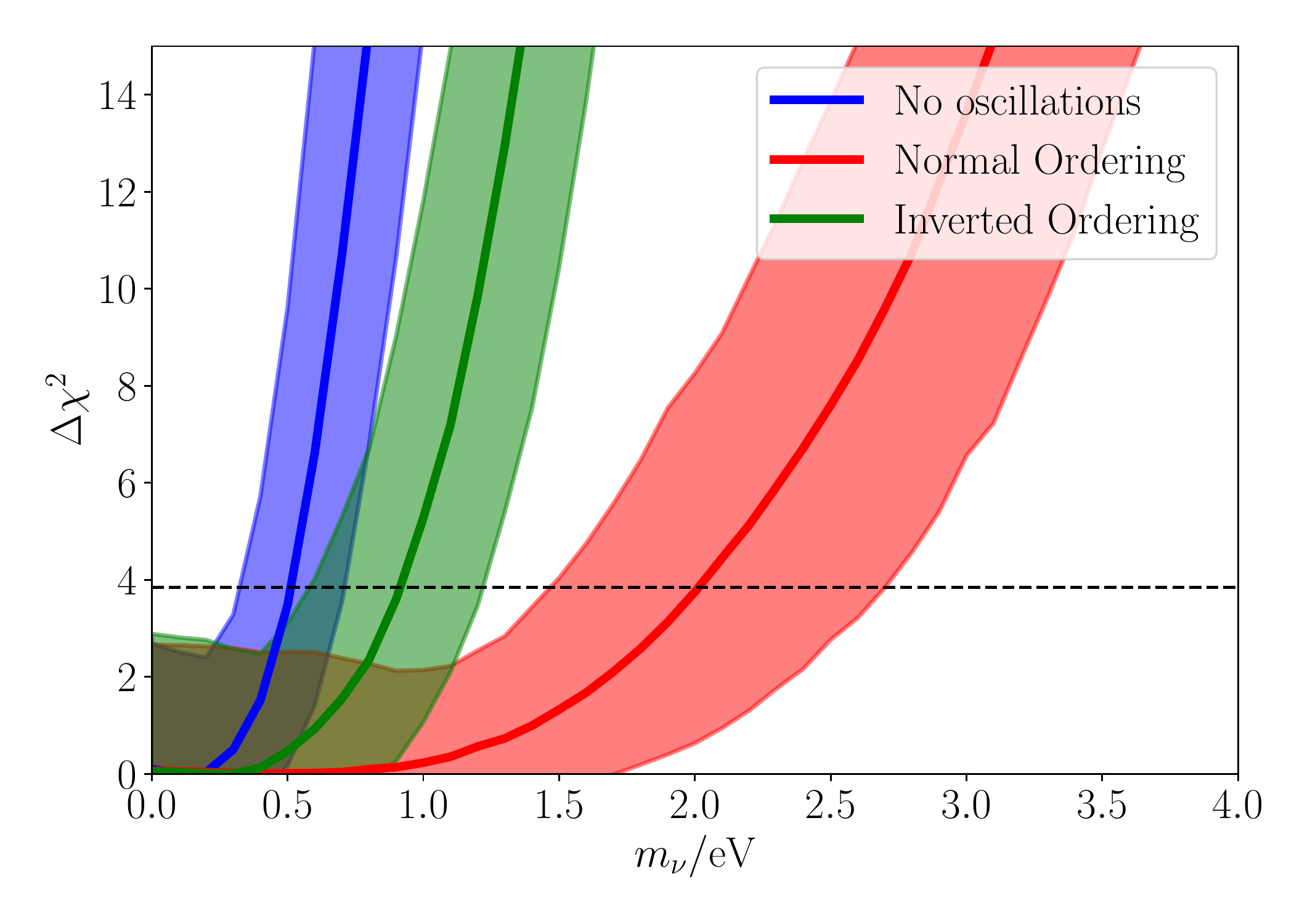}
\caption{\label{fig:chi2}$\Delta\chi^2(m_\nu)$ profiles as a function of neutrino mass $m_\nu$, for DUNE generated samples assuming massless neutrinos and a SN distance of 10~kpc. The mean sensitivities and their $\pm 1\sigma$ uncertainties are shown with solid lines and filled bands, respectively. The horizontal dotted line depicts the $95\%$~CL.}
\end{center}
\end{figure}

\begin{table}
\centering
\caption{Mean and standard deviation of the $95\%$~CL sensitivity on neutrino mass from a sample of DUNE SN datasets at $D=10$~kpc, for different neutrino oscillation scenarios. For the IO case, we give sensitivities for different zenith angles $\theta$.}
\label{tab:m_nu_mass_bounds}
\begin{tabular}{@{\extracolsep{0.5cm}}ccc@{\extracolsep{0cm}}}
\toprule
Neutrino mass ordering  & $\cos\theta$ & $m_\nu$(eV) \\
\midrule
No oscillations                    & $0$ & $0.51^{+0.20}_{-0.20}$ \\
\midrule
Normal Ordering                    & $0$ & $2.01^{+0.69}_{-0.55}$ \\
\midrule
\multirow{5}*{Inverted Ordering}   & $0$ & $0.91^{+0.31}_{-0.33}$ \\
                                   & $-0.5$ & $0.88^{+0.29}_{-0.33}$ \\
                                   & $-1$ & $0.87^{+0.32}_{-0.28}$ \\
\bottomrule 
\end{tabular}
\end{table}

Our statistical procedure, and its results for a SN distance of $D=10$~kpc, can be seen in Fig.~\ref{fig:chi2}. The $\Delta\chi^2$ profiles as a function of neutrino mass are shown for no oscillations, and oscillations in SN environment assuming either NO or IO. Earth matter effects have been neglected in all cases. After including Earth matter effects as previously described, only the IO expectation is affected. Table~\ref{tab:m_nu_mass_bounds} reports our results on the mean and standard deviation of the $m_{\nu}$ sensitivity values for different $\cos\theta$ values, that is, for different angular locations of the SN with respect to the Earth.

As can be seen from Fig.~\ref{fig:chi2} and Tab.~\ref{tab:m_nu_mass_bounds}, 95\% CL sensitivities in the 0.5--2.0~eV range are expected. The best results are expected for the no oscillations and IO scenarios, where the reach is below 1~eV. Despite the largest overall event statistics, $R=1372$, the NO reach is the worst among the three cases, of order 2.0~eV. This result clearly indicates the importance of the shape information, in particular of the sharp neutronization burst time structure visible in Fig.~\ref{fig:events} only for the no oscillations and IO cases. Table~\ref{tab:m_nu_mass_bounds} also shows that oscillations in the Earth's interior barely affect the neutrino mass sensitivity. 

\begin{figure}
\begin{center}
\includegraphics[width=\columnwidth]{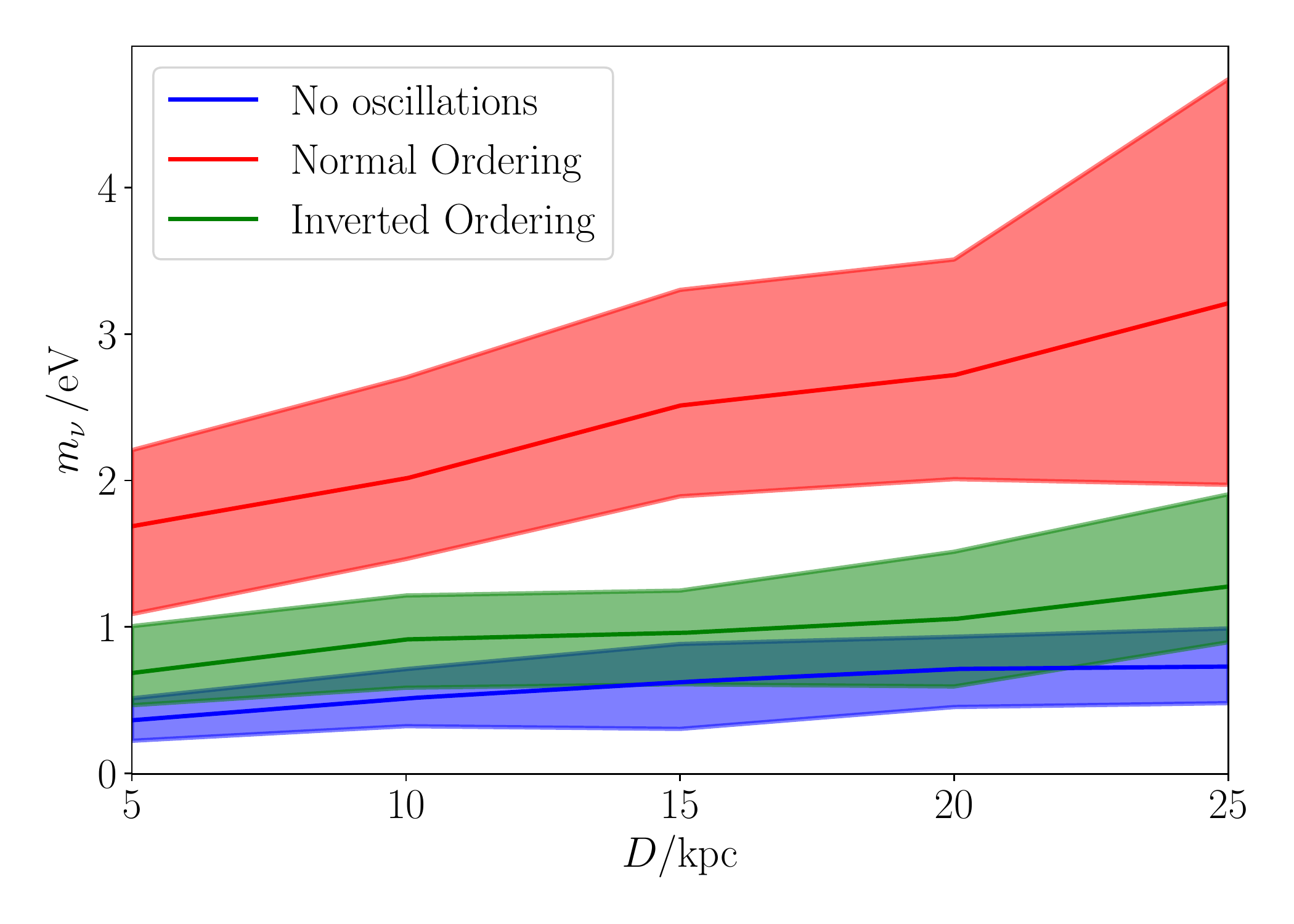} 
\caption{\label{fig:distance}Dependence of the $95\%$~CL neutrino mass sensitivity with the distance $D$ from Earth at which the SN explodes. The mean and standard deviation of the expected sensitivity values are shown with solid lines and filled bands, respectively.}
\end{center}
\end{figure}

Figure ~\ref{fig:distance} shows how the $95\%~$CL sensitivity on the neutrino mass varies with the SN distance $D$. Both the mean and standard deviation of the expected sensitivity values are shown. In all scenarios, the sensitivities to $m_\nu$ worsen by about a factor of 2 as the SN distance increases from 5 to 25~kpc. As is well known, as the distance $D$ increases, the reduced event rate ($R\propto 1/D^2$) tends to be compensated by the increased time delays for a given $m_\nu$ ($\Delta t_i(m_\nu)\propto D$). Our analysis shows that this compensation is only partial, and better sensitivities are obtained for nearby SNe.
A remark is in order. The sensitivity to $m_\nu$ presented so far refers to a low mass progenitor of 8.8$M_\odot$. 
A more massive progenitor usually produces a higher number of events during the accretion and cooling phase \cite{Janka:2017vlw}, whereas no significant change is expected in the neutronization burst, which is a nearly progenitor independent feature \cite{Kachelriess:2004ds}. Therefore, with larger masses, the results reported in Tab. \ref{tab:m_nu_mass_bounds}  for inverted ordering and no oscillations do not change, whereas they can be significantly improved in normal ordering, since in this case the sensitivity depends on the statistics collected in the entire $\sim10$ seconds of the emission.

\section{Conclusions} \label{sec:conclusions}

The capability to detect the electron neutrino flux component from a core-collapse SN in our galactic neighborhood makes large liquid argon detectors powerful observatories to obtain constraints on the absolute value of neutrino mass via time of flight measurements.
Exploiting the signal coming from charged-current interactions of $\nu_e$ with argon nuclei, a 0.9~eV sensitivity on the absolute value of neutrino mass has been obtained in DUNE for IO of neutrino masses, a SN distance of 10~kpc and at 95\% CL. The sensitivity is expected to be significantly worse in NO scenario, 2.0~eV for the same SN distance and confidence level. The sensitivity difference between the two orderings demonstrates the benefit of detecting the $\nu_e$ neutronization burst, whose sharp time structure would be almost entirely suppressed in NO while it should be clearly observable in DUNE if the mass ordering is IO. 
Earth matter induced oscillations, mildly affecting only the IO, and dependence on the SN distance from Earth, have both been studied. The DUNE sensitivity reach appears to be competitive with both laboratory-based direct neutrino mass experiments (such as KATRIN) and next-generation SN observatories primarily sensitive to the $\bar{\nu}_e$ flux component (such as Hyper-Kamiokande and JUNO).

\begin{acknowledgments}
The authors would like to thank John Beacom and Adam Burrows for their helpful comments. This work has been supported by the MCIN/AEI/10.13039/501100011033 of Spain under grant PID2020-113644GB-I00, by the Generalitat Valenciana of Spain under grants PROMETEO/2019/083, PROMETEO/2021/087 and CDEIGENT/2020/003, and by the European Union’s Framework Programme for Research and Innovation Horizon 2020 (2014–2020) under grant H2020-MSCA-ITN-2019/860881-HIDDeN.
\end{acknowledgments}

\bibliography{biblio}

\newpage 
\onecolumngrid
\begin{center}
 \bf \large Supplemental Material
\end{center}
\vspace*{0.2cm}

\twocolumngrid
Here  we  provide  additional  details concerning the expected backgrounds in DUNE, as well as the impact of uncertainties on neutrino production, propagation and interactions.

\section{Backgrounds}
The sensitivities shown in Fig.~\ref{fig:chi2} and Tab.~\ref{tab:m_nu_mass_bounds} of the manuscript can be affected by the presence of cosmogenic and radiogenic backgrounds, as well as neutral current interactions with $^{40}$Ar and elastic scattering on electrons of supernova neutrinos. 
We discuss briefly why this is not the case. 
As shown in \cite{Zhu:2018rwc}, the cosmogenic background at the DUNE far detector underground location is completely negligible, given the very low muon rates ($\simeq$0.05~Hz per module). Radiogenic background is dominated by $(\alpha,n)$ reactions originating from uranium and thorium contamination in the rock surrounding the detector. Neutrons are then captured by $^{40}$Ar, and the subsequent de-excitation of $^{41}$Ar yields a cascade of gamma-rays depositing 6.1~MeV of energy per capture in the argon. This process has a much larger rate, $\sim 5$ events with $E>$5~MeV per second and per detector module \cite{Zhu:2018rwc}. However, this rate can be drastically reduced either via a moderate increase in the energy threshold, or via modest shielding. Concerning the energy threshold, we note that the radiogenic background is concentrated below $E\sim7$ MeV (electron energy) even after considering a 10\% energy resolution, thus becoming completely negligible if we increase our analysis' electron energy threshold from 5 to 7~MeV. We have verified that DUNE's neutrino mass sensitivity is barely affected if such a higher electron energy threshold is adopted, as expected from the fact that only a small fraction ($\simeq$ 5\%) of SN interactions detected in DUNE have electron energies in the range $5<E<7$~MeV. Concerning shielding, the radiogenic background for $E>5$~MeV can also be reduced by two orders of magnitude via a passive neutron absorber of only 20~cm water equivalent thickness \cite{Zhu:2018rwc}, which may be feasible at least for selected (low-background) far detector modules \cite{Avasthi:2022tjr}. 

Despite theoretical predictions of the neutrino neutral current cross-section may differ by one order of magnitude \cite{Capozzi:2018dat}, the corresponding number of events produced cannot be larger than 10\% of the statistics from charged current interactions. This channel has basically no energy information and thus energy reconstruction of charged current events can be worsened, but we checked that doubling the width of the resolution function does not introduce any significant change in the sensitivity. Finally, elastic scattering  produces $O(100)$ events at $D=10$~kpc. In this case, the energy of the neutrino can be partially reconstructed, so further sensitivity to $m_\nu$ might be added. On the other hand, considering conservatively this channel as a background, its statistics can be significantly cut by more than 80\% by performing an angular cut (see \cite{Capozzi:2018dat} where this is discussed in the context of solar neutrinos).

\begin{figure*}
\begin{center}
\includegraphics[width=0.49\textwidth]{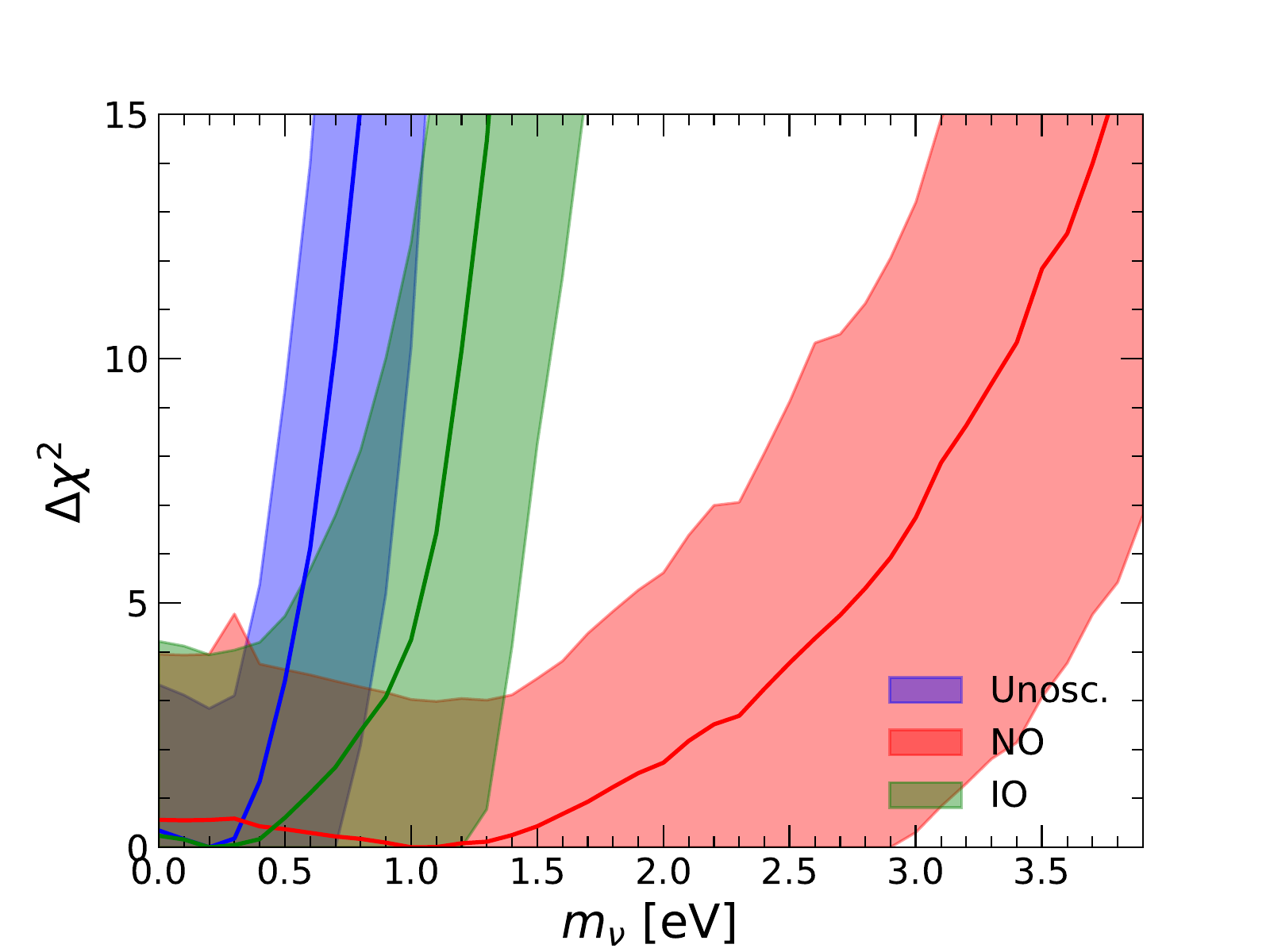} \hfill
\includegraphics[width=0.49\textwidth]{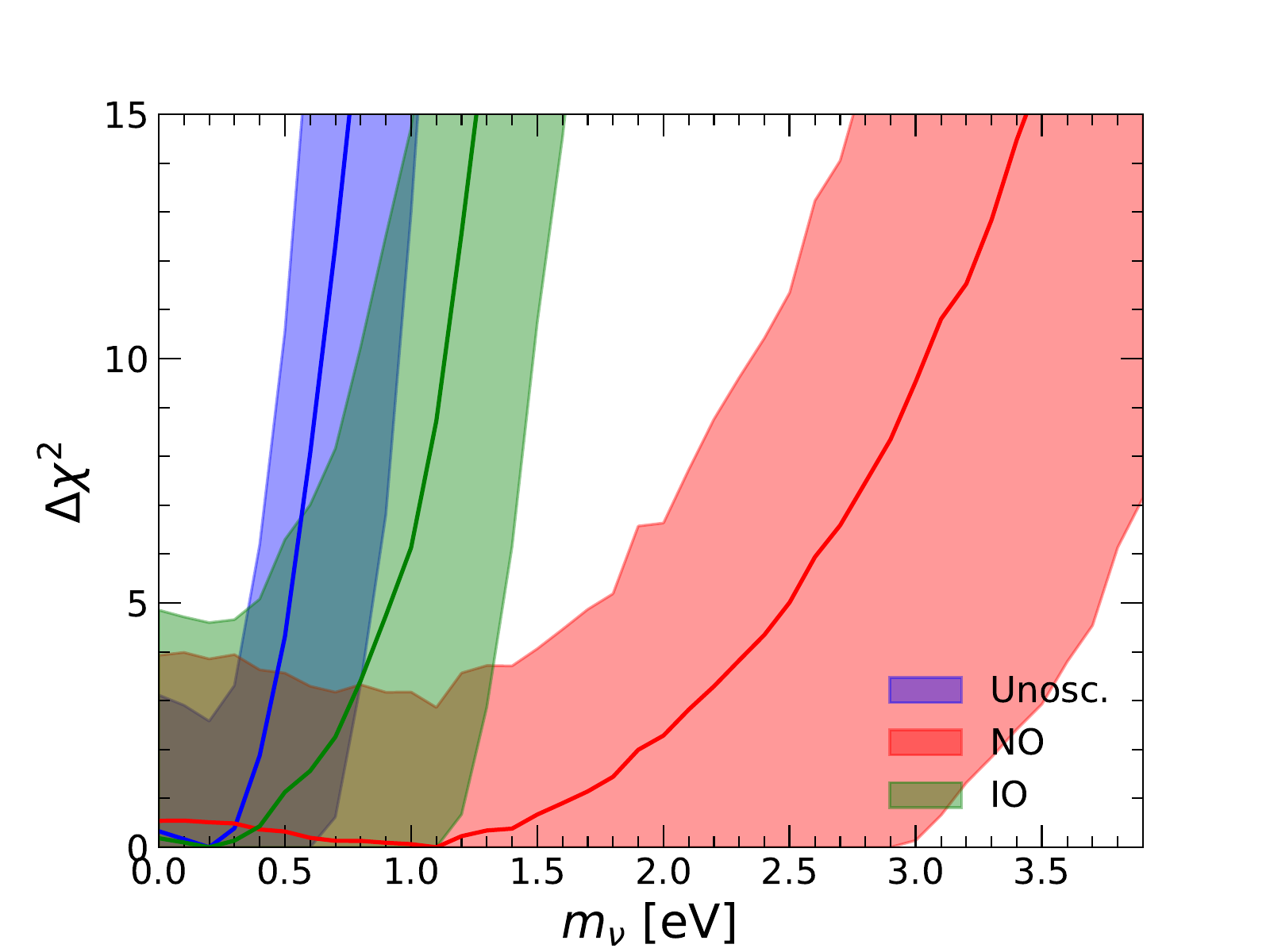}
\caption{\label{fig:sensitivity_check}(Left) Same as Fig.~2 in the main text, but with two extra free parameters $f^{1,2}_{\nu_\alpha}$ which modify the supernova parameters in the following way: $\langle E_{\nu_\alpha}\rangle=(1+f^1_{\nu_\alpha})\langle E_{\nu_\alpha}\rangle^0$ and $\alpha_{\nu_\alpha}=(1+f^2_{\nu_\alpha})\alpha_{\nu_\alpha}^0$, where the superscript 0 refers to the values predicted by the supernova model. We do not set any external (Gaussian) prior on $f^{1,2}$. (Right) Same as the left panel, but with independent free parameters $f_{1,2}$ in two time windows $[0,0.5]$ s and $[0.5,10]$ s.}
\end{center}
\end{figure*}

\section{Systematic uncertainties}
The sensitivity to $m_\nu$ can also be affected by the theoretical uncertainties in neutrino production, propagation and interactions. 

\subsection{Production}
A supernova model can be described as a function of the parameters $\langle E_{\nu_\alpha}\rangle$, $L_{\nu_\alpha}$ and $\alpha_{\nu_\alpha}$ as a function of time. 
This prediction comes unavoidably with an uncertainty. 
In particular, being the time delay proportional to $m_\nu/E_\nu$, $\langle E_{\nu_\alpha}\rangle$ and $\alpha_{\nu_\alpha}$ are somewhat degenerate with $m_\nu$ and their uncertainties can reduce the sensitivity to the neutrino mass. 
Being such uncertainties not easy to assess, we extend our benchmark analysis by taking two extra free parameters $f^{1,2}_{\nu_\alpha}$, which modify the supernova parameters in the following way: $\langle E_{\nu_\alpha}\rangle=(1+f^1_{\nu_\alpha})\langle E_{\nu_\alpha}\rangle^0$ and $\alpha_{\nu_\alpha}=(1+f^2_{\nu_\alpha})\alpha_{\nu_\alpha}^0$, where the superscript 0 refers to the values predicted by the supernova model. 
We do not set any external (gaussian) prior on $f^{1,2}$. 
We also considered a case in which we double the number of free parameters by considering two sets of independent $f^{1,2}_{\nu_\alpha}$ in two different time windows: one in $[0,0.5]$ s (accretion phase) and one in $[0.5,10]$ s (cooling phase). 
We keep using the same likelihood of Eq.~\ref{eq:likelihood_fun}. 
This means that we only consider information on the shape of the signal and not on the total normalization. 
Figure~\ref{fig:sensitivity_check} shows the sensitivity to $m_\nu$ in the cases with time-independent and time-dependent $f^{1,2}_{\nu_\alpha}$, respectively. 
We do not observe significant modifications of the sensitivities to $m_\nu$ reported in Tab.~\ref{tab:m_nu_mass_bounds}. The reason for the overall stability of the sensitivity is that the effect of $m_\nu$ is the introduction of a time delay, whereas any change in either $\langle E_{\nu_\alpha}\rangle$ or $\alpha_{\nu_\alpha}$ implies a modification of the energy spectrum. Moreover, the time delay introduced by $m_\nu$ is energy dependent, meaning that low energy events will arrive later compared to high energy ones, and such an effect cannot be mimicked by neither $\langle E_{\nu_\alpha}\rangle$ nor $\alpha_{\nu_\alpha}$.

\subsection{Propagation}
The survival probability of electron neutrinos only depend on the mixing angles $\sin^2\theta_{13}$ and $\sin^2\theta_{12}$ in NO and IO, respectively, since only oscillations in the supernova environment are relevant. Such mixing angles are currently known with a precision of a few percent and, even when taking the values allowed at $3\sigma$, their variations induce a change in the total number of events smaller than the statistical error. Moreover, in our analysis, we only consider shape information and not the overall normalization.

\subsection{Interactions}
There is a $O(10\%)$ error on the charged current cross-section \cite{Capozzi:2018dat}. If such an uncertainty is treated as an overall normalization error, then we do not expect any impact because we only perform a shape analysis. Even if treated as an energy dependent uncertainty, this uncertainty would induce a smaller effect compared to the analysis with $\langle E_{\nu_\alpha}\rangle$ and $\alpha_{\nu_\alpha}$ production parameters as completely free, hence having negligible impact.

\end{document}